\documentstyle[aps,amssymb]{revtex}

\begin{document}
\title{Boundary term, extended Witten identities and positivity of energy }
\author{G. Y. Chee, Jingfei Zhang}
\address{Physics Department, Liaoning Normal University, Dalian, 116029, China}
\author{Yongxin Guo}
\address{Physics Department, Liaoning University, Shenyang, 110036, China}
\maketitle

\begin{abstract}
In terms of two-spinors a chiral formulation of general relativity with the
Ashtekar Lagrangian and its Hamiltonian formalism in which the basic dynamic
variables are the dyad spinors are presented. The extended Witten identities
are derived. A new expression of the Hamiltonian boundary term is obtained.
Using this expression and the extended Witten identities the proof of the
positive energy theorem is extended to a case including momentum and angular
momentum.

\end{abstract}

\section{Introduction}

It is generally agreed that gravitational energy exists, but because of the
equivalence principle it cannot be localized [1]. A suitable expression
which could provide a physical reasonable description of the energy-momentum
density for gravitating system has lone been sought. All candidates had
several shortcoming. In particular they violated a fundamental theoretical
requirement---that gravitational energy should be positive---as well as
requirements concerning localization and reference frame independence.

During the last 30 years one of the greatest achievements in classical
general relativity is certainly the proof of the positivity of the total
gravitational energy, both at spatial and null infinity [2-4]. It is
precisely positivity that makes this notion not only important (because of
its theoretical significance), but a useful tool as well in the everyday
practice of working relativists. The successes in the proof of the
positivity of energy inspired several relativists to search for expressions
of the gravitational energy-momentum at the quasilocal level, i.e. to
associate these physical quantities with closed spacelike 2-surfaces $S$ (
see, for instance, Ref. [5] and the references therein).

Quasilocal energy is important for its several potential applications. It
can be used to define binding energy of stars. In numerical relativity it
can tell us how to cut off the data on a noncompact region to a compact
region where the quasilocal mass of the compact region approximates the ADM
mass of the noncompact region. When we evolve the data according to Einstein
equations, we need local control of energy to see how the space changes. The
positivity of quasilocal energy is essential for such an investigation. This
could be regarded as a generalization of the energy method in the theory of
a nonlinear hyperbolic system.

There are several contenders for a good definition of quasilocal energy. The
two that interest us are Brown and York's 'canonical quasilocal energy'[6],
and the various definitions based on the integral over $S$\ of the
Witten-Nester two-form (the two-form used in Witten's proof of the positive
energy theorem [3]). The approach of Brown and York starts with standard
Hamilton-Jacobi theory and thus is somewhat different from more traditional
approaches based on Noether techniques. Brown and York have determined what
geometric entity play the role of quasilocal energy in general relativity.
The canonical quasilocal energy is supposed to be the energy of the
gravitational and matter fields contained in a spatial volume $\Sigma $,
whose boundary two-surface is $S=\partial \Sigma $.{\em \ } This approach
gives the correct quasilocal energy surface density and allows one the
freedom to assign the reference point of the quasilocal energy.

The approach based on the Witten-Nester two-form uses spinorial methods, and
the different definitions are distinguished by the choice of supplementary
equation the $S$-spinors are supposed to satisfy, for example the Sen-Witten
equation [3], the Dougan-Mason equation [7], or the Ludvigsen-Vickers
equation [8]. Since Witten's positive energy proof can be understood in
terms of the Hamiltonian, the Hamiltonian density associated with this proof
provides a locally positive localization---and thus has real promise as a
truly physical energy-momentum density for gravitational fields. To fulfill
this promise certain features need further consideration; an outstanding one
concerns the role of---and even the need for---the spinor field, which
seemed rather mysterious.{\em \ } Although there are very beautiful argument
for some nice important results, it is not clear as to how much the Witten
argument really captures the correct physics. Moreover, all the spinor
formulations cannot give angular momentum and the center-of-mass moment [9].

The two approaches are conceptually distinct, even though in some
circumstances one might expect their numerical value to coincide. Therefore,
it is of interest to establish a connection between the Brown-York
quasilocal energy and the Witten-Nester form. Indeed, such a connection
supports the use of the Witten-Nerster form in the quasilocal context, and,
furthermore, provides one with a non-spinor vantage point from which the
various supplementary spinor equations can be examined. The general
relationship between the Brown-York quasilocal energy and the common
''spinorial definition of quasilocal energy'' constructed from the
Witten-Nerster two-form has been established by Lau [10]. He found that the
Sen-Witten energy expression can always be written in a form which is
essentially identical to the that of the Brown-York quasilocal energy. His
theory is an Ashtekar-variable reformulation of the metric theory of
quasilocal stress-energy-momentum originally due to Brown and York. It is
meaningful to apply Lau's formalism to the proof of the positive energy
theorem. In this paper we start with the self-dual Lagrangian following
Lau's route and then use Witten-Ashtekar-Horowitz's method [11] to prove the
positive energy theorem. To this end we have to derive a set of extended
Witten identities and then apply them to the positivity proof. In the
discussion on positivity proof and energy-momentum quasilocalization the
boundary term has important roles. The value of the Hamiltonian ''on shell''
is determined by the Hamiltonian boundary term. The ''natural'' Hamiltonian
boundary term inherited from the Lagrangian can---and should---be adjusted.
The general expression for boundary term depends on the choice of variables,
a displacement vector field (e.g. translation for energy momentum and
rotation for angular momentum), a reference configuration and boundary
conditions. Being different from most of the works on positivity of energy,
in this paper by choosing a boundary term with a nonvanishing shift we will
extend the proof to involve momentum and even angular momentum.

The organization of this paper is as follows. In Sec. II. a review of some
results of Lau is given in a rather different appearance. The Lagrangian of
general relativity is written in terms of two-spinors and then is decomposed
into two independent parts, the self-dual and the anti-self-dual part. The
self-dual part is identified with a version of the Ashtekar Lagrangian [12]
given by Goldberg [13] and then can be taken as the gravitational
Lagrangian. Sec. III is the Hamiltonian formulation in which the dyad
spinors are chosen as the basic dynamic variables while the Ashtekar
connections appear in the conjugate momenta. In Sec. IV the Hamiltonian
boundary term is rewritten as a suitable form for the application of the
extended Witten identities. The extended Witten identities are derived in
Sec. V and are used to calculate the boundary term and prove the positivity
of the energy in Sec. VI. Finally, Sec. VII is devoted to some conclusions
and remarks.

\section{The self-dual Lagrangian of GR}

We start by rewriting some results of Lau [10] in a rather different
appearance in order to adapt to our purpose. Using the spin connection
one-form $\omega _a{}^{KL}$ the curvature two-form $R_{abKL}$ of the
spacetime $M$ can be written as [14] 
\begin{equation}
R_{abKL}=\nabla _a^\Gamma \omega _{bKL}-\nabla _b^\Gamma \omega
_{aKL}-\omega _a{}^M{}_K\omega _{bML}+\omega _b{}^M{}_K\omega _{aML},
\end{equation}
where $\nabla _a^\Gamma $ is the covariant derivative defined by the affine
connection $\Gamma ^a{}_{bc}$, for example, the covariant derivative of the
cotetrad one-form $e^J{}_b{}$ is 
\[
\nabla _a^\Gamma e^J{}_b=e^I{}_b\omega _{aI}{}^J=\partial _ae_{Jb}-\Gamma
^c{}_{ab}e^J{}_c. 
\]
Then the scalar curvature $R$ can be written as 
\begin{equation}
R=2\nabla _a^\Gamma \left( e^{Ka}e^{Lb}\omega _{bKL}\right)
+e^{Ka}{}e^{Lb}\omega _a{}^I{}_K\omega _{bIL}-e^{Ka}e^{Lb}{}\omega
_{aL}{}^I\omega _{bKI},
\end{equation}
and the gravitational Lagrangian ${\cal L}_G$ consists of the Moller
Lagrangian [15] and a divergence term: 
\begin{eqnarray*}
{\cal L}_G &=&\sqrt{-g}R \\
&=&\sqrt{-g}\left( e^{Ka}{}e^{Lb}\omega _a{}^I{}_K\omega
_{bIL}-e^{Ka}e^{Lb}{}\omega _{aL}{}^I\omega _{bKI}\right) +2\partial
_a\left( \sqrt{-g}e^{Ka}e^{Lb}\omega _{bKL}\right) .
\end{eqnarray*}

Passing on to two-spinor expression [16], the spin connection reads 
\begin{eqnarray}
\omega _{AA^{\prime }BB^{\prime }}{}^{CC^{\prime }} &=&\omega _{AA^{\prime
}B}{}^C\epsilon _{B^{\prime }}{}^{C^{\prime }}+\overline{\omega }%
_{AA^{\prime }B^{\prime }}{}^{C\prime }\epsilon _B{}^C  \nonumber \\
&=&\omega _{AA^{\prime }BB^{\prime }}^{+}{}^{CC^{\prime }}+\omega
_{AA^{\prime }BB^{\prime }}^{-}{}^{CC^{\prime }},
\end{eqnarray}
where 
\begin{equation}
\omega _{AA^{\prime }BB^{\prime }}^{+}{}^{CC^{\prime }}=\omega _{AA^{\prime
}B}{}^C\epsilon _{B^{\prime }}{}^{C^{\prime }},
\end{equation}
and 
\begin{equation}
\overline{\omega }_{AA^{\prime }BB^{\prime }}{}^{CC^{\prime }}=\overline{%
\omega }_{AA^{\prime }B^{\prime }}{}^{C\prime }\epsilon _B{}^C,
\end{equation}
is the self-dual and the anti-self-dual part of the connection $\omega
_{AA^{\prime }BB^{\prime }}{}^{CC^{\prime }}$, respectively, and then the
scalar curvature $R$ splits into two parts 
\begin{equation}
R=R^{+}+R^{-},
\end{equation}
where the self-dual and the anti-self-dual part of $R$ is, respectively 
\begin{eqnarray}
R^{+} &=&2\nabla _a^\Gamma \left( e^{KK^{\prime }a}e^{LL^{\prime }b}\omega
_{bKL}\epsilon _{L^{\prime }K^{\prime }}\right) +\omega ^K{}_{K^{\prime
}}{}^I{}_K\omega ^{LK^{\prime }}{}_{IL}-\omega {}^{KK^{\prime }I}{}_L\omega
^L{}_{K^{\prime }IK} \\
R^{-} &=&2\nabla _a^\Gamma \left( e^{KK^{\prime }a}e^{LL^{\prime }b}%
\overline{\omega }_{bK^{\prime }L^{\prime }}\epsilon _{LK}\right) +\overline{%
\omega }_K{}^{K^{\prime }I^{\prime }}{}_{K^{\prime }}\overline{\omega }%
^{LL^{\prime }}{}_{I^{\prime }L^{\prime }}-\overline{\omega }{}^{KK^{\prime
}I^{\prime }}{}_{L^{\prime }}\overline{\omega }_K{}^{L^{\prime
}}{}_{I^{\prime }K^{\prime }}.
\end{eqnarray}
Consequently, the gravitational Lagrangian ${\cal L}_G=\sqrt{-g}R$ splits
into two pats 
\[
{\cal L}_G={\cal L}_G^{+}+{\cal L}_G^{-}. 
\]
The self-dual part ${\cal L}_G^{+}$ consists of the self-dual Moller
Lagrangian [13,15] and a divergence term: 
\begin{equation}
{\cal L}_G^{+}=^{(4)}\sigma \left( \omega ^K{}_{K^{\prime }}{}^I{}_K\omega
^{LK^{\prime }}{}_{IL}-\omega {}^{KK^{\prime }I}{}_L\omega ^L{}_{K^{\prime
}IK}\right) +\partial _a\left( 2^{(4)}\sigma e^{KK^{\prime }a}e^{LL^{\prime
}b}\omega _{bKL}\epsilon _{L^{\prime }K^{\prime }}\right)
\end{equation}
where 
\begin{equation}
^{(4)}\sigma =\sqrt{-g}
\end{equation}
is the determinant of the SL(2,C) soldering form $\sigma _\mu {}^{AA^{\prime
}}$ on the spacetime manifold $M$. Since ${\cal L}_G^{+}$ and ${\cal L}%
_G^{-} $ depend on $\omega _{AA^{\prime }B}{}^C$ and $\overline{\omega }%
_{AA^{\prime }B^{\prime }}{}^{C\prime }$ respectively, and are independent
of each other, we can choose the self-dual part ${\cal L}_G^{+}$ as the
Lagrangian of the gravitational field, which is identified with a version of
the Ashtekar Lagrangian [12] given by Goldberg [13].

\section{The Hamiltonian formalism of the self-dual gravity}

To put the theory in Hamiltonian form, we will assume that the spacetime $M$
is topologically $\Sigma \times R$ for some spacelike submanifold $\Sigma $
and assume that there exists a time function $t$ with nowhere vanishing
gradient $\left( dt\right) _a$ such that each $t=$const. surface $\Sigma _t$
is diffeomorphic to $\Sigma $. $t^a$ will denote the time flow vector field
( $t^a\left( dt\right) _a=1$), while $n^a$ will denote the future-pointing
timelike vector field ($n^an_a=-1$) normal to the $t=$const. surfaces. Then
the intrinsic metric $q_{ab}$ of $\Sigma $, the lapse $N$, and the shift $%
N^a $ are induced by 
\[
g_{ab}=q_{ab}-n_an_b, 
\]
and 
\[
t^a=Nn^a+N^a, 
\]
respectively. In the spinor notation [16], the normal vector $n^{AA^{\prime
}}$ defines an isomorphism from the space of primed spinors to the space of
unprimed spinors: 
\[
\alpha ^{+A}=i\sqrt{2}n^{AA^{\prime }}\overline{\alpha }_{A^{\prime }}, 
\]
\[
\omega _{ABCD}=i\sqrt{2}n_B{}^{A^{\prime }}\omega _{AA^{\prime }CD}, 
\]
\begin{equation}
n^{AB}=i\sqrt{2}n^{BA^{\prime }}n^A{}_{A^{\prime }}=\frac i{\sqrt{2}}%
\epsilon ^{AB}.
\end{equation}
In this notation 
\[
g_{ab}=q_{ab}-n_an_b, 
\]
reads 
\[
g^{ABCD}=q^{ABCD}-n^{AB}n^{CD}, 
\]
or 
\begin{equation}
\epsilon ^{AC}\epsilon ^{BD}=-\epsilon ^{A(C}\epsilon ^{D)B}+\frac 12%
\epsilon ^{AB}\epsilon ^{CD}.
\end{equation}
In terms of the dyad spinors 
\[
\zeta _0{}^C=o^C,\zeta _1{}^C=\iota ^C, 
\]
the self-dual connection can be expressed as 
\begin{equation}
\omega _{ABC}{}^D=\zeta _C{}^b\triangledown _{AB}^\Gamma \zeta _b{}^D,
\end{equation}
Using these results we can decompose the quantities appearing in ${\cal L}%
_G^{+}$ into the tangential and the normal parts to the surface $\Sigma _t$,
for example, for a four-vector $V^{AB}$ we have 
\begin{eqnarray*}
V^{AB} &=&q_{CD}^{AB}V^{CD}-n^{AB}n_{CD}V^{CD} \\
&=&V^{\left( AB\right) }+V^{\left[ AB\right] },
\end{eqnarray*}
where $V^{\left( AB\right) }$ and $V^{\left[ AB\right] }$ is the tangential
and the normal part of $V^{AB}$ to the surface $\Sigma _t$, separately. Then
the Lagrangian ${\cal L}_G^{+}$ can be rewritten as 
\begin{eqnarray}
{\cal L}_G^{+} &=&4\sqrt{2}i\sigma \nabla ^{\Gamma \left( AB\right) }\zeta
^a{}_B\stackrel{\cdot }{\zeta }_{aA}  \nonumber \\
&&-N\sigma [\triangledown _{\left( AB\right) }^\Gamma \zeta ^A{}_a\nabla
^{\Gamma \left( CB\right) }\zeta _C{}^a-\triangledown _{\left( AB\right)
}^\Gamma \zeta _C{}^a\nabla ^{\Gamma \left( BC\right) }\zeta ^A{}_a] 
\nonumber \\
&&-\sqrt{2}i\sigma N^{CD}\left( 3\nabla _{CD}^\Gamma \zeta
_A{}^a\triangledown ^{\Gamma \left( AB\right) }\zeta _{aB}+\zeta
_A{}^a\nabla _{CD}^\Gamma \triangledown ^{\Gamma \left( AB\right) }\zeta
_{aB}\right)  \nonumber \\
&&-\partial _{\left( AB\right) }\left( 2N\sigma \zeta {}^{Ab}\triangledown
^{\Gamma \left( CB\right) }\zeta _{bC}-\sqrt{2}i\sigma N^{CD}\zeta
^{Aa}\nabla _{\left( CD\right) }^\Gamma \zeta _a{}^B\right) ,
\end{eqnarray}
where 
\begin{equation}
\stackrel{\cdot }{\zeta }_{aA}=t^{CD}\triangledown _{CD}^\Gamma \zeta _{aA},
\end{equation}
and 
\begin{equation}
\sigma =\frac{^4\sigma }N=\frac{\sqrt{-g}}N.
\end{equation}
The dyad spinors $\zeta _{aA}$ are chosen as the basic dynamic variables and
then the canonical momenta conjugate to $\zeta _{aA}$ are 
\begin{equation}
\widetilde{p}^{aA}=\frac{\partial {\cal L}_G^{+}}{\partial \stackrel{.}{%
\zeta }_{aA}}=4\sqrt{2}i\sigma {}\nabla ^{\Gamma \left( AB\right) }\zeta
^a{}_B.
\end{equation}
The gravitational Hamiltonian density can be computed 
\begin{eqnarray}
{\cal H}_G &=&\widetilde{p}^{aA}\stackrel{.}{\zeta }_{aA}-{\cal L}_G^{+} \\
&=&N\sigma [\triangledown _{\left( AB\right) }^\Gamma \zeta ^A{}_a\nabla
^{\Gamma \left( CB\right) }\zeta _C{}^a-\triangledown _{\left( AB\right)
}^\Gamma \zeta _C{}^a\nabla ^{\Gamma \left( BC\right) }\zeta ^A{}_a] 
\nonumber \\
&&+\sqrt{2}i\sigma N^{CD}\left[ 3\left( \nabla _{CD}^\Gamma \zeta
_A{}^a\right) \triangledown ^{\Gamma \left( AB\right) }\zeta _{aB}+\zeta
_A{}^b\nabla _{CD}^\Gamma \triangledown ^{\Gamma \left( AB\right) }\zeta
_{bB}\right]  \nonumber \\
&&+\partial _{\left( AB\right) }\left( 2N\sigma \zeta {}^{Ab}\triangledown
^{\Gamma \left( CB\right) }\zeta _{bC}-\sqrt{2}i\sigma N^{CD}\zeta
^{Aa}\nabla _{\left( CD\right) }^\Gamma \zeta _a{}^B\right) .
\end{eqnarray}
Using 
\[
\nabla ^{\Gamma \left( AB\right) }\zeta ^a{}_B=-\frac i{4\sqrt{2}\sigma {}}%
\widetilde{p}^{aA}, 
\]
we obtain 
\begin{eqnarray}
{\cal H}_G &=&N[\frac 1{32\sigma {}}\widetilde{p}_{aA}\widetilde{p}%
^{aA}-\sigma \triangledown _{\left( AB\right) }^\Gamma \zeta _C{}^a\nabla
^{\Gamma \left( BC\right) }\zeta ^A{}_a]  \nonumber \\
&&-\frac 1{4{}}N^{CD}\left[ 3\left( \nabla _{CD}^\Gamma \zeta _{aA}{}\right) 
\widetilde{p}^{aA}+\sigma \zeta _{aA}{}^b\nabla _{CD}^\Gamma \left( \frac 1{%
\sigma {}}\widetilde{p}^{aA}\right) \right]  \nonumber \\
&&+\partial _{\left( AB\right) }\left( \frac{\sqrt{2}i}{4{}}N\zeta _a{}^A%
\widetilde{p}^{aB}-\sqrt{2}i\sigma N^{CD}\zeta ^{Aa}\nabla _{\left(
CD\right) }^\Gamma \zeta _a{}^B\right)
\end{eqnarray}
i. e. 
\begin{equation}
{\cal H}_G=N{\cal H}+N^{AB}{\cal H}_{AB}+\partial _{\left( AB\right) }%
\widetilde{B}{}^{AB},
\end{equation}
where the Hamiltonian constraint 
\begin{equation}
{\cal H}=\frac 1{32{}\sigma }\widetilde{p}_{aA}\widetilde{p}^{aA}-\sigma
\triangledown _{\left( AB\right) }^\Gamma \zeta _{aC}{}\nabla ^{\Gamma
\left( BC\right) }\zeta ^{aA},
\end{equation}
the momentum constraint 
\begin{equation}
{\cal H}_{AB}=-\frac 34\widetilde{p}^{aC}\nabla _{AB}^\Gamma \zeta _{aC}{}-%
\frac 14\sigma \zeta _{aC}{}\nabla _{\left( AB\right) }^\Gamma \left( \frac 1%
\sigma \widetilde{p}^{aC}\right) ,
\end{equation}
and the boundary term 
\begin{eqnarray}
\widetilde{B}{}^{AB} &=&2N\sigma \zeta {}^{Ab}\triangledown ^{\Gamma \left(
CB\right) }\zeta _{bC}-\sqrt{2}i\sigma N^{CD}\zeta ^{Aa}\nabla _{\left(
CD\right) }^\Gamma \zeta _a{}^B  \nonumber \\
&=&\frac{\sqrt{2}i}{4{}}N\zeta _a{}^A\widetilde{p}^{aB}-\sqrt{2}i\sigma
N^{CD}\zeta ^{Aa}\nabla _{\left( CD\right) }^\Gamma \zeta _a{}^B,
\end{eqnarray}
separately. Being different from Ashtekar's formulation, here the dyad
spinors $\zeta _{aA}$ are chosen as the basic dynamic variables and the
conjugate momenta $\widetilde{p}^{aA}$ are related to Ashtekar's variables $%
\omega _{\left( AB\right) {}}{}^{CD}$ [12,13] by 
\[
{}\omega _{\left( AC\right) B}{}^C=\frac i{4\sqrt{2}\sigma }\zeta _B{}^a%
\widetilde{p}_{aA}{}. 
\]

\section{Boundary terms}

Being different from most of the works on positivity of energy in which the
shift $N^{AB}$ is chosen to vanish, we consider a more general boundary term 
$\widetilde{B}{}^{AB}$ which allows a particular choice of $N^{AB}$.
Choosing the triad 
\begin{eqnarray}
\sigma _1{}^{AB} &=&\frac i{\sqrt{2}}\left( o^Ao^B-\iota ^A\iota ^B\right) =%
\frac 1{\sqrt{2}}\left( m^a+\overline{m}^a\right) ,  \nonumber \\
\sigma _2{}^{AB} &=&\frac 1{\sqrt{2}}\left( o^Ao^B+\iota ^A\iota ^B\right) =-%
\frac i{\sqrt{2}}\left( m^a-\overline{m}^a\right) ,  \nonumber \\
\sigma _3{}^{AB} &=&\frac i{\sqrt{2}}\left( -o^A\iota ^B-\iota ^Ao^B\right) =%
\frac 1{\sqrt{2}}\left( l^a-n^a\right) ,
\end{eqnarray}
on $\Sigma $ then the normal unit vector to the boundary $\partial \Sigma $
is 
\[
v{}^{AB}=\sigma _3{}^{AB}=-\frac i{\sqrt{2}}\left( o^A\iota ^B+\iota
^Ao^B\right) . 
\]
We chose 
\begin{eqnarray}
N &=&-\chi ^2, \\
N^{AB} &=&\chi ^2\left( m^{AB}+\overline{m}^{AB}+v^{AB}\right)  \nonumber \\
&=&\frac i{\sqrt{2}}\chi ^2\left( o^Ao^B+\iota ^A\iota ^B-o^{(A}\iota
^{B)}\right) ,
\end{eqnarray}
and compute the boundary integral 
\begin{eqnarray}
\oint \widetilde{B}{}^{AB}dS_{AB} &=&\oint \widetilde{B}{}^{AB}v_{AB}dS \\
&=&-\sqrt{2}\oint \sigma \chi ^2\left( \overline{m}^bm_a\nabla _bl^a-m^b%
\overline{m}_a\nabla _bn^a\right) dS  \nonumber \\
&&-\frac{\sqrt{2}}2\oint \sigma \chi ^2\left( \overline{m}^b-m^b-\frac{{}3}2%
l^b+\frac 12n^b\right) \left( n_a\nabla _bl^a-\overline{m}_a\nabla
_bm^a\right) dS  \nonumber \\
&=&-\oint \left[ \sigma \chi ^2k-\frac{\sqrt{2}}2\sigma \chi ^2\left( 
\overline{m}^b-m^b-\frac{{}3}2l^b+\frac 12n^b\right) \left( n_a\nabla _bl^a-%
\overline{m}_a\nabla _bm^a\right) \right] dS
\end{eqnarray}
where $k$ is the trace of the extrinsic curvature of $\partial \Sigma $. The
first term of the integral is just {\em {\bf \ }}the ''unrefernced''
Brown-York quasilocal energy when $\chi ^2=1$[6,10], the second term comes
from the particular choice (27) of $N^{AB}$ and then includes the
contribution of momentum and angular momentum on which our interest is
concentrated.

For further calculation and discussion we rewritten the expression of $%
\widetilde{B}{}^{AB}$ as follows.

Since 
\begin{eqnarray*}
&&\partial _{\left( AB\right) }\left( \sqrt{2}i\sigma N^{CD}\omega
_{CD}{}^{AB}\right) \\
&=&\sqrt{2}i\sigma D_{\left( AB\right) }^{\left( 3\right) }\left(
N^{CD}\omega _{CD}{}^{AB}\right) \\
&=&\sqrt{2}i\sigma \epsilon ^{AE}\epsilon ^{BF}\epsilon ^{CG}\epsilon
^{DH}D_{\left( EF\right) }^{\left( 3\right) }\left( N_{GH}\omega
_{CDAB}{}\right) \\
&=&\sqrt{2}i\sigma \left( \epsilon ^{DB}\epsilon ^{FH}+\epsilon
^{FD}\epsilon ^{BH}\right) \left( \epsilon ^{CA}\epsilon ^{EG}+\epsilon
^{EC}\epsilon ^{AG}\right) D_{\left( EF\right) }^{\left( 3\right) }\left(
N_{GH}\omega _{CDAB}{}\right) \\
&=&\sqrt{2}i\sigma D_{\left( AB\right) }^{\left( 3\right) }\left(
N^{AB}\omega _{CD}{}^{CD}\right) -\sqrt{2}i\sigma D_{\left( AB\right)
}^{\left( 3\right) }\left( N{}^{AD}\omega _C{}^{BC}{}_D\right) \\
&&-\sqrt{2}i\sigma D_{\left( AB\right) }^{\left( 3\right) }\left(
N{}^{CA}\omega {}^{BD}{}_{CD}\right) +\sqrt{2}i\sigma D_{\left( AB\right)
}^{\left( 3\right) }\left( N^{CD}\omega {}^{AB}{}_{CD}\right) \\
&=&\sqrt{2}i\partial _{\left( AB\right) }\left( \sigma N^{AB}\zeta
{}^{Cb}\triangledown _{\left( CD\right) }^\Gamma \zeta _b{}^D+\sigma
N^{CD}\zeta _C{}^b\triangledown ^{\Gamma AB}\zeta _{bD}{}\right) ,
\end{eqnarray*}
$\widetilde{B}{}^{AB}$ can be written as 
\begin{eqnarray}
\widetilde{B}{}^{AB} &=&2N\sigma \zeta {}^{Aa}\triangledown ^{\Gamma \left(
CB\right) }\zeta _{aC}-\sqrt{2}i\sigma N^{AB}\zeta {}^{Cb}\triangledown
_{\left( CD\right) }^\Gamma \zeta _b{}^D  \nonumber \\
&&-\sqrt{2}i\sigma N^{CD}\zeta _C{}^a\triangledown ^{\Gamma \left( AB\right)
}\zeta _{aD}.
\end{eqnarray}

Introducing spinors $\lambda ^A$ and $\lambda ^{+A}$ by 
\begin{equation}
o^A=\frac 1\chi \lambda ^A,\iota ^A=\frac 1\chi \lambda ^{+A}
\end{equation}
and noting 
\begin{equation}
\nabla _{AA^{\prime }}^\Gamma \lambda {}^C=\partial _{AA^{\prime }}\lambda
{}^C,
\end{equation}
we can compute 
\begin{eqnarray*}
\widetilde{B}{}^{AB} &=&2N\sigma \frac 1{\chi ^2}\left( \lambda ^A\partial
^{\left( CB\right) }\lambda _C^{+}-\lambda ^{+A}\partial ^{\left( CB\right)
}\lambda _C\right) +2N\sigma \frac 1{\chi ^3}\left( \lambda ^{+A}\lambda
_C-\lambda ^A\lambda _C^{+}\right) \partial ^{\left( CB\right) }\chi \\
&&-\sqrt{2}i\sigma N^{AB}\frac 1{\chi ^2}\left( \lambda {}^C\partial
_{\left( CD\right) }\lambda {}^{+D}-\lambda {}^{+C}\partial _{\left(
CD\right) }\lambda {}^D\right) \\
&&-\sqrt{2}i\sigma N^{CD}\frac 1{\chi ^2}\left( \lambda _C{}\partial
^{\left( AB\right) }\lambda _D^{+}-\lambda _C^{+}{}\partial ^{\left(
AB\right) }\lambda _D\right) .
\end{eqnarray*}
Noting 
\begin{equation}
N=-\chi ^2=-\lambda _A\lambda ^{+A},
\end{equation}
and 
\begin{eqnarray}
N^{AB} &=&-m^{AB}-\overline{m}^{AB}-v^{AB}  \nonumber \\
&=&\frac i{\sqrt{2}}\left( \lambda ^A\lambda ^B+\lambda ^{+A}\lambda
^{+B}+\lambda ^{(A}\lambda ^{+B)}\right) ,
\end{eqnarray}
we obtain 
\begin{eqnarray}
\widetilde{B}{}^{AB} &=&-\frac \sigma {\chi ^2}\left( \lambda ^A\lambda
^B+\lambda ^{+A}\lambda ^{+B}+\lambda ^{(A}\lambda ^{+B)}\right) \left(
\lambda {}^C\partial _{\left( CD\right) }\lambda {}^{+D}-\lambda
{}^{+C}\partial _{\left( CD\right) }\lambda {}^D\right)  \nonumber \\
&&-2\sigma \left( \lambda ^A\partial ^{\left( CB\right) }\lambda
_C^{+}-\lambda ^{+A}\partial ^{\left( CB\right) }\lambda _C\right) -\sigma
\left( \lambda ^{+A}\lambda _C-\lambda ^A\lambda _C^{+}\right) \partial
^{\left( CB\right) }\ln \chi ^2  \nonumber \\
&&+\sigma \left( \lambda ^{+D}{}\partial ^{\left( AB\right) }\lambda
_D^{+}+\lambda ^D{}\partial ^{\left( AB\right) }\lambda _D+\frac 12\lambda
^D{}\partial ^{\left( AB\right) }\lambda _D^{+}+\frac 12\lambda
^{+D}{}\partial ^{\left( AB\right) }\lambda _D\right) .
\end{eqnarray}

\section{The extended Witten identity}

In order to deal with the boundary term $\widetilde{B}{}^{AB}$ we follow
Witten [3], Ashtekar and Horowitz [11] and derive some useful identities. In
a two-spinor formulation a Dirac spinor $\varepsilon $ and Dirac matrices $%
\gamma ^I$ can be expressed 
\[
\varepsilon =\left( 
\begin{array}{l}
\lambda ^A \\ 
\overline{\tau }_{A^{\prime }}
\end{array}
\right) ,\gamma ^I=\sqrt{2}\left( 
\begin{array}{ll}
0 & \sigma ^{IAA^{\prime }} \\ 
\sigma ^I{}_{A^{\prime }A} & 0
\end{array}
\right) , 
\]
respectively, and then the Witten identity [3] 
\begin{equation}
-\nabla ^2\varepsilon =(i\nabla {\cal )}^2\varepsilon =-\sum_i^i\nabla
^i\nabla _i\varepsilon +4\pi G(T_{00}+\sum_jT_{0j}\gamma ^0\gamma
^j)\varepsilon .
\end{equation}
becomes two two-spinor identities 
\begin{eqnarray}
-2\nabla ^{AC}\nabla _{CB}\lambda ^B &=&-\nabla ^{BC}\nabla _{BC}\lambda
^A+4\pi G(T_{00}\lambda ^A+\sqrt{2}T_0{}^A{}_B\lambda ^B), \\
-2\overline{\nabla }_{A^{\prime }C^{\prime }}\overline{\nabla }^{C^{\prime
}B^{\prime }}\overline{\tau }_{B^{\prime }} &=&-\overline{\nabla }%
^{B^{\prime }C^{\prime }}\overline{\nabla }_{B^{\prime }C^{\prime }}%
\overline{\tau }_{A^{\prime }}+4\pi G(T_{00}\overline{\tau }_{A^{\prime }}+%
\sqrt{2}T_{0A^{\prime }}{}^{B^{\prime }}\overline{\tau }_{B^{\prime }}), 
\nonumber
\end{eqnarray}
where $\nabla _{AB}$ and $\overline{\nabla }_{A^{\prime }B^{\prime }}$
denote the covariant derivatives defined by the spin connections $\omega
_{ABC}{}^D$ and $\overline{\omega }_{A^{\prime }B^{\prime }C^{\prime
}}{}^{D^{\prime }}$ , respectively. Multiplying the first identity with $%
\lambda ^{+A}$ leads to 
\[
2\lambda ^{+A}\nabla _{\left( AC\right) }\nabla ^{\left( CB\right) }\lambda
_B=\lambda ^{+A}\nabla _{\left( BC\right) }\nabla ^{\left( BC\right)
}\lambda _A-4\pi G(T_{00}\lambda ^{+A}\lambda _A+\sqrt{2}T_{0AB}{}\lambda
^{+A}\lambda ^B). 
\]
Since 
\begin{eqnarray*}
\lambda ^{+A}\nabla _{\left( BC\right) }\nabla ^{\left( BC\right) }\lambda
_A &=&\nabla _{\left( BC\right) }\left( \lambda ^{+A}\nabla ^{\left(
BC\right) }\lambda _A\right) -\left( \nabla _{\left( BC\right) }\lambda
^{+A}\right) \nabla ^{\left( BC\right) }\lambda _A \\
&=&D_{BC}\left( \lambda ^{+A}\nabla ^{\left( BC\right) }\lambda _A\right)
-\left( \nabla _{\left( BC\right) }\lambda ^A\right) ^{+}\nabla ^{\left(
BC\right) }\lambda _A,
\end{eqnarray*}
then we have 
\begin{eqnarray*}
\sigma D_{BC}\left( \lambda ^{+A}\nabla ^{\left( BC\right) }\lambda
_A\right) &=&2\sigma \lambda ^{+A}\nabla _{\left( AC\right) }\nabla ^{\left(
CB\right) }\lambda _B+\sigma \left( \nabla _{\left( BC\right) }\lambda
^A\right) ^{+}\nabla ^{\left( BC\right) }\lambda _A \\
&&+4\pi G\sigma (T_{00}\lambda ^{+A}\lambda _A+\sqrt{2}T_{0AB}{}\lambda
^{+A}\lambda ^B).
\end{eqnarray*}
Integrating it leads to 
\begin{eqnarray}
\oint \sigma \lambda ^{+A}\nabla ^{\left( BC\right) }\lambda _AdS_{\left(
AB\right) } &=&\int_\Sigma \sigma D_{BC}\left( \lambda ^{+A}\nabla ^{\left(
BC\right) }\lambda _A\right) dV  \nonumber \\
&=&2\int_\Sigma \sigma \lambda ^{+A}\nabla _{\left( AC\right) }\nabla
^{\left( CB\right) }\lambda _BdV+\int_\Sigma \sigma \left( \nabla _{\left(
BC\right) }\lambda ^A\right) ^{+}\nabla ^{\left( BC\right) }\lambda _AdV 
\nonumber \\
&&+4\pi G\int_\Sigma \sigma (T_{00}\lambda ^{+A}\lambda _A+\sqrt{2}%
T_{0AB}{}\lambda ^{+A}\lambda ^B)dV.
\end{eqnarray}
Multiplying the first identity of (37) with $\lambda ^A$ leads to 
\[
D_{BC}\left( \lambda ^A\nabla ^{\left( BC\right) }\lambda _A\right)
=2\lambda ^A\nabla _{\left( AC\right) }\nabla ^{\left( CB\right) }\lambda
_B+4\sqrt{2}\pi GT_{0AB}{}\lambda ^A\lambda ^B. 
\]
Integrating yields 
\begin{eqnarray}
\oint \sigma \lambda ^A\nabla ^{\left( BC\right) }\lambda _AdS_{\left(
AB\right) } &=&\int_\Sigma \sigma D_{BC}\left( \lambda ^A\nabla ^{\left(
BC\right) }\lambda _A\right) dV  \nonumber \\
&=&2\int_\Sigma \sigma \lambda ^A\nabla _{\left( AC\right) }\nabla ^{\left(
CB\right) }\lambda _BdV+4\sqrt{2}\pi G\int_\Sigma \sigma T_{0AB}{}\lambda
^A\lambda ^BdV.
\end{eqnarray}

By the similar way, from the second identity of (37) we obtain 
\begin{eqnarray}
\oint \sigma \lambda ^{+A}\nabla _a\lambda _A^{+}dS^a &=&2\int_\Sigma \sigma 
\overline{\lambda }^{A^{\prime }}\overline{\nabla }_{\left( A^{\prime
}C^{\prime }\right) }\overline{\nabla }^{\left( C^{\prime }B^{\prime
}\right) }\overline{\lambda }_{B^{\prime }}dV+4\sqrt{2}\pi G\int_\Sigma
\sigma \overline{\lambda }^{A^{\prime }}T_{0A^{\prime }}{}^{B^{\prime }}%
\overline{\lambda }_{B^{\prime }}dV  \nonumber \\
&&-i\sqrt{2}\oint \sigma \lambda ^{+B}K_{CD}{}^A{}_B\lambda _A^{+}dS^{\left(
BC\right) }
\end{eqnarray}
and 
\begin{eqnarray}
&&\oint \sigma \lambda ^A\nabla _a\lambda _A^{+}dS^a  \nonumber \\
&=&2\int_\Sigma \sigma \lambda ^{A^{\prime }}\overline{\nabla }_{\left(
A^{\prime }C^{\prime }\right) }\overline{\nabla }^{\left( C^{\prime
}B^{\prime }\right) }\overline{\lambda }_{B^{\prime }}dV+\int_\Sigma \nabla
_{\left( BC\right) }\lambda ^A\left( \nabla ^{\left( BC\right) }\lambda
_A\right) ^{+}dV  \nonumber \\
&&+4\pi G\int_\Sigma \sigma (T_{00}\lambda ^{A^{\prime }}\overline{\lambda }%
_{A^{\prime }}+\sqrt{2}\lambda ^{A^{\prime }}T_{0A^{\prime }}{}^{B^{\prime }}%
\overline{\lambda }_{B^{\prime }})dV+i\sqrt{2}\oint \sigma \lambda
^BK_{CD}{}^A{}_B\lambda _A^{+}dS^{\left( CD\right) }  \nonumber \\
&&-i\sqrt{2}\int_\Sigma \sigma K_{CDA}{}^B\left( \lambda ^A\nabla ^{\left(
CD\right) }\lambda _B^{+}-\lambda _B^{+}\nabla ^{\left( CD\right) }\lambda
^A\right) dV+2\int_\Sigma \sigma K_{CDA}{}^EK^{CDB}{}_E\lambda ^A\lambda
_B^{+}dV.
\end{eqnarray}

(38), (39),(40), and (41) can be called the {\em extended Witten identities}.

\section{The calculation of the boundary integral}

Now we are ready to calculate the boundary integral. Since

\[
\zeta _{0A}=o_A=\frac 1\chi \lambda _A,\zeta _{1A}=\iota _A=\frac 1\chi
\lambda _A^{+}, 
\]

\[
\nabla _{AA^{\prime }}^\Gamma \lambda {}^C=\partial _{AA^{\prime }}\lambda
{}^C, 
\]
then 
\begin{eqnarray*}
\omega _{AA^{\prime }B}{}^C &=&\zeta _B{}^b\triangledown _{AA^{\prime
}}^\Gamma \zeta _b{}^C \\
&=&\frac 1{\chi ^2}\left( \lambda _B{}\partial _{AA^{\prime }}\lambda
^{+C}-\lambda _B^{+}{}\partial _{AA^{\prime }}\lambda {}^C\right) +\frac 1{%
\chi ^3}\partial _{AA^{\prime }}\chi \left( \lambda _B^{+}{}\lambda
{}^C-\lambda _B{}\lambda ^{+C}\right) .
\end{eqnarray*}
and 
\begin{eqnarray*}
\nabla _{AA^{\prime }}\lambda _B &:&=\partial _{AA^{\prime }}\lambda
_B-\omega _{AA^{\prime }B}{}^C\lambda _C \\
&=&\partial _{AA^{\prime }}\lambda _B-\frac 1{\chi ^2}\left( \lambda
_B{}\partial _{AA^{\prime }}\lambda ^{+C}-\lambda _B^{+}{}\partial
_{AA^{\prime }}\lambda {}^C\right) \lambda _C+\frac 1\chi \partial
_{AA^{\prime }}\chi \lambda _B{}.
\end{eqnarray*}
Using this result we obtain 
\[
\lambda ^{+B}\nabla _{AA^{\prime }}\lambda _B=2\lambda ^{+B}\partial
_{AA^{\prime }}\lambda _B-\chi \partial _{AA^{\prime }}\chi , 
\]
and 
\[
\lambda ^B\nabla _{AA^{\prime }}\lambda _B=\lambda ^B\partial _{AA^{\prime
}}\lambda _B-{}\partial _{AA^{\prime }}\lambda {}^C\lambda _C=2\lambda
^B\partial _{AA^{\prime }}\lambda _B. 
\]
By the same way we find that 
\begin{eqnarray*}
\nabla _{AA^{\prime }}\lambda _B^{+} &=&\partial _{AA^{\prime }}\lambda
_B^{+}-\omega _{AA^{\prime }B}{}^C\lambda _C^{+} \\
&=&\partial _{AA^{\prime }}\lambda _B^{+}-\frac 1{\chi ^2}\left( \lambda
_B{}\partial _{AA^{\prime }}\lambda ^{+C}-\lambda _B^{+}{}\partial
_{AA^{\prime }}\lambda {}^C\right) \lambda _C^{+}+\frac 1\chi \partial
_{AA^{\prime }}\chi \lambda _B^{+}{},
\end{eqnarray*}
which leads to 
\[
\lambda ^B\nabla _{AA^{\prime }}\lambda _B^{+}=2\lambda ^B\partial
_{AA^{\prime }}\lambda _B^{+}+\chi \partial _{AA^{\prime }}\chi , 
\]
and 
\[
\lambda ^{+B}\nabla _{AA^{\prime }}\lambda _B^{+}=2\lambda ^{+B}\partial
_{AA^{\prime }}\lambda _B^{+}. 
\]
Then we have 
\begin{eqnarray*}
\lambda ^{+B}\partial _{AA^{\prime }}\lambda _B+\lambda ^B\partial
_{AA^{\prime }}\lambda _B^{+} &=&\frac 12\left( \lambda ^{+B}\nabla
_{AA^{\prime }}\lambda _B+\lambda ^B\nabla _{AA^{\prime }}\lambda
_B^{+}\right) , \\
\lambda ^B\partial _{AA^{\prime }}\lambda _B &=&\frac 12\lambda ^B\nabla
_{AA^{\prime }}\lambda _B, \\
\lambda ^{+B}\partial _{AA^{\prime }}\lambda _B^{+} &=&\frac 12\lambda
^{+B}\nabla _{AA^{\prime }}\lambda _B^{+},
\end{eqnarray*}
and 
\begin{eqnarray*}
\lambda ^{+D}\partial ^{\left( AB\right) }\lambda _D+\lambda ^D\partial
^{\left( AB\right) }\lambda _D^{+} &=&\frac 12\left( \lambda ^{+D}\nabla
^{\left( AB\right) }\lambda _D+\lambda ^D\nabla ^{\left( AB\right) }\lambda
_D^{+}\right) , \\
\lambda ^D\partial ^{\left( AB\right) }\lambda _D &=&\frac 12\lambda
^D\nabla ^{\left( AB\right) }\lambda _D, \\
\lambda ^{+D}\partial ^{\left( AB\right) }\lambda _D^{+} &=&\frac 12\lambda
^{+D}\nabla ^{\left( AB\right) }\lambda _D^{+}.
\end{eqnarray*}
Using these results and (35) we obtain 
\begin{eqnarray*}
\oint \widetilde{B}{}^{AB}dS_{\left( AB\right) } &=&-\oint \frac \sigma {%
\chi ^2}\left( \lambda ^A\lambda ^B+\lambda ^{+A}\lambda ^{+B}+\lambda
^{(A}\lambda ^{+B)}\right) \left( \lambda {}^C\partial _{\left( CD\right)
}\lambda {}^{+D}-\lambda {}^{+C}\partial _{\left( CD\right) }\lambda
{}^D\right) dS_{\left( AB\right) } \\
&&-2\oint \sigma \left( \lambda ^A\partial ^{\left( CB\right) }\lambda
_C^{+}-\lambda ^{+A}\partial ^{\left( CB\right) }\lambda _C\right)
dS_{\left( AB\right) } \\
&&-\oint \sigma \left[ \frac 2\chi \left( \lambda ^{+A}\lambda _C-\lambda
^A\lambda _C^{+}\right) \partial ^{\left( CB\right) }\chi +\frac \chi 2%
\partial ^{\left( AB\right) }\chi \right] dS_{\left( AB\right) } \\
&&+\frac 12\oint \sigma \left( \lambda ^{+D}{}\nabla \partial ^{\left(
AB\right) }\lambda _D^{+}+\lambda ^D{}\nabla ^{\left( AB\right) }\lambda
_D+\lambda ^{+D}\nabla ^{\left( AB\right) }\lambda _D\right) dS_{\left(
AB\right) }
\end{eqnarray*}
From the extended Witten identities (38), (39), and (40) one finds 
\begin{eqnarray*}
&&\frac 12\oint \sigma \left( \lambda ^{+D}{}\nabla ^{\left( AB\right)
}\lambda _D^{+}+\lambda ^D{}\nabla ^{\left( AB\right) }\lambda _D+\lambda
^{+D}{}\nabla ^{\left( AB\right) }\lambda _D\right) dS_{\left( AB\right) } \\
&=&\int_\Sigma \sigma \lambda ^{+A}\nabla _{\left( AC\right) }\nabla
^{\left( CB\right) }\lambda _BdV+\int_\Sigma \sigma \lambda ^A\nabla
_{\left( AC\right) }\nabla ^{\left( CB\right) }\lambda _BdV+\int_\Sigma
\sigma \overline{\lambda }^{A^{\prime }}\overline{\nabla }_{\left( A^{\prime
}C^{\prime }\right) }\overline{\nabla }^{\left( C^{\prime }B^{\prime
}\right) }\overline{\lambda }_{B^{\prime }}dV \\
&&+\frac 12\int_\Sigma \sigma \left( \nabla _{\left( BC\right) }\lambda
^A\right) ^{+}\nabla ^{\left( BC\right) }\lambda _AdV+2\pi G\int_\Sigma
\sigma (T_{00}\lambda ^{+A}\lambda _A+\sqrt{2}T_{0AB}{}\lambda ^{+A}\lambda
^B)dV \\
&&-\frac{i\sqrt{2}}2\oint \sigma \lambda ^{+B}K{}^{CDA}{}_B\lambda
_A^{+}dS_{\left( CD\right) },
\end{eqnarray*}
and then the boundary integral becomes 
\begin{eqnarray}
&&\oint \widetilde{B}{}^{AB}dS_{\left( BC\right) }  \nonumber \\
&=&-\oint \frac \sigma {\chi ^2}\left( \lambda ^A\lambda ^B+\lambda
^{+A}\lambda ^{+B}+\lambda ^{(A}\lambda ^{+B)}\right) \left( \lambda
{}^C\partial _{\left( CD\right) }\lambda {}^{+D}-\lambda {}^{+C}\partial
_{\left( CD\right) }\lambda {}^D\right) dS_{\left( AB\right) }  \nonumber \\
&&-2\oint \sigma \left( \lambda ^A\partial ^{\left( CB\right) }\lambda
_C^{+}-\lambda ^{+A}\partial ^{\left( CB\right) }\lambda _C\right)
dS_{\left( BC\right) }  \nonumber \\
&&-\oint \sigma \left[ \frac 2\chi \left( \lambda ^{+A}\lambda _C-\lambda
^A\lambda _C^{+}\right) \partial ^{\left( CB\right) }\chi +\frac 12\chi
\partial ^{\left( AB\right) }\chi \right] dS_{\left( BC\right) }  \nonumber
\\
&&+\int_\Sigma \sigma \lambda ^{+A}\nabla _{\left( AC\right) }\nabla
^{\left( CB\right) }\lambda _BdV+\int_\Sigma \sigma \lambda ^A\nabla
_{\left( AC\right) }\nabla ^{\left( CB\right) }\lambda _BdV+\int_\Sigma
\sigma \overline{\lambda }^{A^{\prime }}\overline{\nabla }_{\left( A^{\prime
}C^{\prime }\right) }\overline{\nabla }^{\left( C^{\prime }B^{\prime
}\right) }\overline{\lambda }_{B^{\prime }}dV  \nonumber \\
&&+\frac 12\int_\Sigma \sigma \left( \nabla _{\left( BC\right) }\lambda
^A\right) ^{+}\nabla ^{\left( BC\right) }\lambda _AdV+2\pi G\int_\Sigma
\sigma (T_{00}\lambda ^{+A}\lambda _A+\sqrt{2}T_{0AB}{}\lambda ^{+A}\lambda
^B)dV  \nonumber \\
&&-\frac{i\sqrt{2}}2\oint \sigma \lambda ^{+B}K{}^{CDA}{}_B\lambda
_A^{+}dS_{\left( CD\right) }.
\end{eqnarray}

When the Witten equations 
\[
\nabla ^{\left( CB\right) }\lambda _B=0,\overline{\nabla }^{\left( C^{\prime
}B^{\prime }\right) }\overline{\lambda }_{B^{\prime }}=0, 
\]
are satisfied we have

\begin{eqnarray}
&&\oint \widetilde{B}{}^{AB}dS_{\left( AB\right) }  \nonumber \\
&=&-\oint \frac \sigma {\chi ^2}\left( \lambda ^A\lambda ^B+\lambda
^{+A}\lambda ^{+B}+\lambda ^{(A}\lambda ^{+B)}\right) \left( \lambda
{}^C\partial _{\left( CD\right) }\lambda {}^{+D}-\lambda {}^{+C}\partial
_{\left( CD\right) }\lambda {}^D\right) dS_{\left( AB\right) }  \nonumber \\
&&-2\oint \sigma \left( \lambda ^A\partial ^{\left( CB\right) }\lambda
_C^{+}-\lambda ^{+A}\partial ^{\left( CB\right) }\lambda _C\right)
dS_{\left( AB\right) }  \nonumber \\
&&-\oint \sigma \left[ \frac 2\chi \left( \lambda ^{+A}\lambda _C-\lambda
^A\lambda _C^{+}\right) \partial ^{\left( CB\right) }\chi +\frac 12\chi
\partial ^{\left( AB\right) }\chi \right] dS_{\left( AB\right) }  \nonumber
\\
&&+\frac 12\int_\Sigma \sigma \left( \nabla _{\left( BC\right) }\lambda
^A\right) ^{+}\nabla ^{\left( BC\right) }\lambda _AdV+2\pi G\int_\Sigma
\sigma (T_{00}\lambda ^{+A}\lambda _A+\sqrt{2}T_{0AB}{}\lambda ^{+A}\lambda
^B)dV  \nonumber \\
&&-\frac{i\sqrt{2}}2\oint \sigma \lambda ^{+B}K{}^{CDA}{}_B\lambda
_A^{+}dS_{\left( CD\right) }.
\end{eqnarray}
Noting 
\[
dS_{\left( AB\right) }=v_{\left( AB\right) }dS, 
\]
and 
\[
v{}^{AB}=-\frac i{\sqrt{2}}\left( o^A\iota ^B+\iota ^Ao^B\right) , 
\]
one finds 
\begin{eqnarray*}
&&-\oint \frac \sigma {\chi ^2}\left( \lambda ^A\lambda ^B+\lambda
^{+A}\lambda ^{+B}+\lambda ^{(A}\lambda ^{+B)}\right) \left( \lambda
{}^C\partial _{\left( CD\right) }\lambda {}^{+D}-\lambda {}^{+C}\partial
_{\left( CD\right) }\lambda {}^D\right) dS_{\left( AB\right) } \\
&=&-\frac i{\sqrt{2}}\oint \sigma \left( \lambda {}^C\partial _{\left(
CD\right) }\lambda {}^{+D}-\lambda {}^{+C}\partial _{\left( CD\right)
}\lambda {}^D\right) dS.
\end{eqnarray*}
And according to the definition of the extrinsic curvature of $\Sigma $ 
\begin{eqnarray*}
K_{CDAB} &=&-\nabla _{\left( CD\right) }\sigma _{0AB} \\
&=&-\frac i{\sqrt{2}}\nabla _{\left( CD\right) }\left( -o^A\iota ^B+\iota
^Ao^B\right) ,
\end{eqnarray*}
we have 
\[
\lambda ^{+B}K{}_{CDAB}\lambda ^{+A}=0. 
\]
Finally, the boundary integral becomes 
\begin{eqnarray}
&&\oint \widetilde{B}{}^{AB}dS_{\left( AB\right) }=-\frac i{\sqrt{2}}\oint
\sigma \left( \lambda {}^C\partial _{\left( CD\right) }\lambda
{}^{+D}-\lambda {}^{+C}\partial _{\left( CD\right) }\lambda {}^D\right) dS 
\nonumber \\
&&-2\oint \sigma \left( \lambda ^A\partial ^{\left( CB\right) }\lambda
_C^{+}-\lambda ^{+A}\partial ^{\left( CB\right) }\lambda _C\right) v_{\left(
AB\right) }dS  \nonumber \\
&&-\oint \sigma \left[ \frac 2\chi \left( \lambda ^{+A}\lambda _C-\lambda
^A\lambda _C^{+}\right) \partial ^{\left( CB\right) }\chi +\frac 12\chi
\partial ^{\left( AB\right) }\chi \right] v_{\left( AB\right) }dS  \nonumber
\\
&&+\frac 12\int_\Sigma \sigma \left( \nabla _{\left( BC\right) }\lambda
^A\right) ^{+}\nabla ^{\left( BC\right) }\lambda _AdV  \nonumber \\
&&+2\pi G\int_\Sigma \sigma (T_{00}\lambda ^{+A}\lambda _A+\sqrt{2}%
T_{0AB}{}\lambda ^{+A}\lambda ^B)dV.
\end{eqnarray}
This is manifestly non-negative if the gravitational source satisfies the
dominant energy condition and then the positivity of the quasilocal energy
can be proved easily. Starting from the choice (34) of the shift vector $%
N^{AB}$, the prove can be extended to include momentum and angular momentum.

\section{Conclusions and remarks}

Using a spinor method the Einstein-Hilbert Lagrangian is decomposed into the
self-dual and the anti-self-dual part. The self-dual part is identified with
a version of the Ashtekar Lagrangian given by Goldberg. Starting from this
Lagrangian and choosing the dyad spinors as the basic dynamical variables a
Hamiltonian formulation is developed. A set of extended Witten identities is
derived. Choosing a particular shift vector $N^{AB}$ and applying the
extended Witten identities the proof of the positive energy theorem is
extended to a case including momentum and angular momentum.

In both the approach of Brown-York and the approach of Witten-Nester the
boundary term has important roles. The ''natural'' Hamiltonian boundary term
inherited from the Lagrangian can---and should---be adjusted. The choice of
the Hamiltonian boundary term is the key point in the discussion of this
paper. Being different from most of the works on positivity of energy in
which the shift $N^{AB}$ is chosen to vanish, we consider a more general
boundary term $\widetilde{B}{}^{AB}$ which includes the 'unreferenced'
Brown-York quasilocal energy and allows a particular choice of $N^{AB}$.
However, we do not deal with the issue of reference point since it has been
shown [10] that when the Witten-Nester expression is evaluated on solution
spinors to the Sen-Witten equation (obeying appropriate boundary
conditions), an implicit reference point for the energy is set. The further
investigation of the boundary term in some particular cases will be given in
a forthcoming paper elsewhere.

ACKNOWLEDGMENT

This work was supported by the National Science Foundation of China Grant
No. 10175032.

\end{document}